\documentclass[final,1p,times]{elsarticle} 

\usepackage{graphicx}
\usepackage{amssymb} 
\usepackage{amsthm} 
\usepackage{lineno}

\journal{Nuclear Physics A} 
 
\begin{document}

\begin{frontmatter} 

\title{Systematic investigation of partonic collectivity through
  centrality dependence of elliptic flow of multi-strange hadrons in
  Au+Au collisions at $\sqrt{s_{NN}}$ = 200  GeV in STAR}

\author{Md. Nasim ( For the STAR collaboration)}
\address{Variable Energy Cyclotron Centre, Kolkata, India}


\begin{abstract} 
   We present the measurement of centrality dependence of
multi-strange hadron ($\Omega$, $\Xi$ and $\phi$) elliptic flow ($v_{2}$) at mid-rapidity in Au+Au collision
at $\sqrt{s_{NN}}$ = 200 GeV. We investigate number-of-constituents 
quark (NCQ) scaling of identified hadrons $v_{2}$ for different
collision centralities. Maximum deviation from ideal NCQ scaling
observed at the level of 10$\%$ for centrality
30-80$\%$ at  $(m_{T} - m_{0})$/$n_{q}$ $>$ 0.6 GeV/$c^{2}$   . This may indicate smaller contribution to
the collectivity from the partonic phase in the peripheral
collisions. We also observe the  mass ordering of $v_{2}$ break down between proton and
$\phi$-meson  for $p_{T}$ $< $1 GeV/$c$ . This could be 
due to the effect of  later stage hadronic interactions on  $v_{2}$.
\end{abstract} 

\end{frontmatter} 


\section{Introduction}

In the non-central nucleus nucleus collisions, the overlapping area is
not spatially isotropic. This initial spatial anisotropy is then transformed into
momentum anisotropy because of pressure gradients developed due to the
subsequent interactions among the constituents~\cite{flow1}. The elliptic flow
($v_{2}$) is  a measure of the anisotropy in momentum space. The
elliptic flow parameter is defined as the second Fourier coefficient
of the particles distribution in emission azimuthal angle ($\phi$)
with respect to the reaction plane angle ($\Psi$) and can be defined
as 
\begin{equation}
v_{2}=\langle\cos(2(\phi-\Psi))\rangle.
\end{equation}
One of the main goals of the STAR experiment at Relativistic Heavy
Ion Collider (RHIC) is to study the properties of the QCD (Quantum
Chromodynamics)
matter at extremely high energy and parton densities,
created in the heavy-ion collisions.
In such a case, measurement of $v_{2}$ plays a crucial role. Within
a hydrodynamical framework $v_{2}$ is an early time phenomenon and
sensitive to the equation of state of the system formed in the
collisions~\cite{flow1}. Thus $v_{2}$ can be used as probe for early system
although its magnitude may change due to later stage hadronic
interactions. The hadronic
interaction cross sections of multi-strange hadrons ($\Omega$, $\Xi$
and $\phi$) is small
and also they freeze-out close to the quark-hadron transition temperature
predicted by lattice QCD~\cite{white,smallX}.  Hence, the multi-strange hadrons
are expected to provide information from the partonic stage
of the evolution in the heavy-ion collisions. Furthermore, the
multi-strange hadron anisotropic flow in heavy-ion collisions when
compared to those from  $K_{S}^{0}$ and $\Lambda$, single strange
valence quark carrying hadrons, will be useful for understanding the
collective dynamics of the strange quarks.
Recent data shows when $v_{2}$ of hadrons are scaled by the number of 
constituent-quarks ($ n_{q} $)  and measure as function of $(m_{ T} -m_{0})/n_{q}$,  where 
$m_{T}$  is the transverse mass and $m_{0}$ is the mass of the hadron, the $v_{2}$ values follow a universal 
scaling for all the measured hadrons and nuclei~\cite{rhicflow}. This observation,
referred to as the number of constituent quark scaling, has been
consider as signature of partonic collectivity in heavy-ion
collisions~\cite{voloshin,starphiflow}.


\section{Data Sets and Methods}

The results presented here are based on high statistics data set collected at
$\sqrt{s_{NN}}$= 200 GeV in Au+Au collisions with the
STAR detector for a minimum bias trigger~\cite{trigger} in the year of 2010.  The total number of
minimum bias events analyzed are about 240
million. The Time
Projection Chamber (TPC)
and Time of Flight (TOF) detectors 
with full $2\pi$ coverage were used for particle identification in the
central rapidity ($\it{y}$) region ($|\it{y}|<$ 1.0).
Particles are identified from
information of the
specific energy loss as a function of momentum (using TPC) and 
square of the mass as a function of momentum (using TOF). We reconstruct
$\Omega^{\pm}$, $\Xi^{\pm}$ and $\phi$ through their decay channel : $\Omega^{-}$
$\longrightarrow$ $\Lambda$ + $\it{K}^{-}$ ($\overline{\Omega}^{+}$$\longrightarrow$ $\overline{\Lambda}$ + $\it{K}^{+}$) , $\Xi^{-}$ $\longrightarrow$ $\Lambda$ +
$\pi^{-}$ ($\overline{\Xi}^{+}$ $\longrightarrow$ $\overline{\Lambda}$ +
$\pi^{+}$) and $\phi$ $\longrightarrow$ $\it{K}^{+}$ + $\it{K}^{-}$. Topological cuts
and kinematic cuts were applied to reduced the combinatorial
background for $\Omega$ and $\Xi$. \\
The $\eta$-sub event plane method~\cite{method} has been used for the
flow analysis. 
In this method, one defines the event flow vector for each
particle based on particles measured in the opposite hemisphere in
pseudo-rapidity ($\eta$).
An $\eta$ gap of $|\eta| <$ 0.05 between positive and negative
pseudo-rapidity sub-events has been introduced to suppress non-flow
effects.
$v_{2}$ vs. invariant mass method has been used to extract $v_{2}$  of
multi-strange hadrons. Details of this method can be found in the
reference~\cite{inv_mass_method}.

\section{Results}
\begin{figure}[!ht]
\centerline{\includegraphics[scale=0.34]{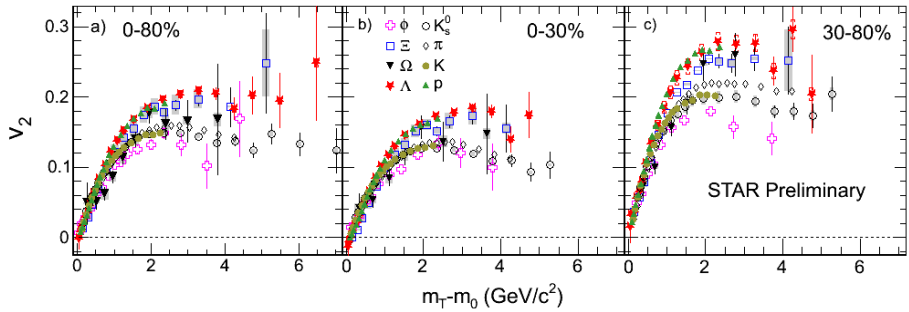}}
\caption{(Color online) $v_{2}$ vs. $m_{ T}-m_{0}$ of identified hadrons for Au + Au collision at
   $\sqrt{s_{NN}}$= 200 GeV with $|\it{y}| < $ 1.0 for centrality a)
   0-80$\%$ b) 0-30$\%$ and c) 30-80$\%$. Systematic errors on
   $K^{0}_{S}$, $\Xi$, and $\Omega$ are shown by shaded band and for
   $\Lambda$ by cap symbol. The vertical lines are the statistical errors.  }
\end{figure} 
The $v_{2}$ of identified hadrons ($\pi$, $\it{K}$,
$\it{p}$, $K^{0}_{S}$, $\Lambda$, $\Xi$, $\Omega$ and $\phi$)   as a function of
transverse kinetic energy $m_{ T}-m_{0}$ at $\sqrt{s_{NN}}$= 200 GeV  for various
centralities are
shown in figure 1.  We observed a  splitting between baryon and
meson $v_{2}$ at intermediate $m_{T}-m_{0}$ for centrality 0-30$\%$. However
for 30-80$\%$ centrality, we observed no such distinct grouping among
the baryons and among the
mesons. For  mesons and baryons, multi-strange hadrons show smaller $v_{2}$ than
 other identified hadrons ($v_{2}$($\phi$) $<$
$v_{2}$($K^{0}_{S}$), $v_{2}$($\Xi$) $<$ $v_{2}$($\Lambda$) )  at
intermediate $m_{ T}-m_{0}$. As  the multi-strange hadrons
$v_{2}$  mostly reflect collectivity from partonic phase, therefore our observation may indicate smaller contribution to the 
collectivity from the partonic phase in the peripheral collisions. 

\begin{figure}[!ht]
\centerline{\includegraphics[scale=0.33]{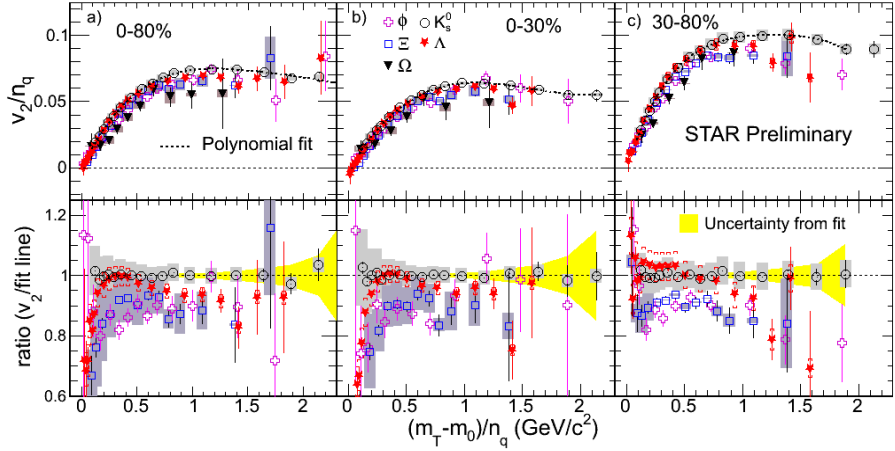}}
\caption{(Color online) $v_{2}/n_{q}$ vs. $(m_{T} -m_{0})/n_{q}$ in Au + Au
collisions at $\sqrt{s_{NN}}$= 200 GeV  $|\it{y}| < $ 1.0 for centrality a)
   0-80 $\%$ b) 0-30 $\%$ and c) 30-80 $\%$. Systematic errors on
   $K^{0}_{S}$, $\Xi$, and $\Omega$ are shown by shaded band and for
   $\Lambda$ by cap symbol. The vertical lines are the statistical
   errors. Dotted black line is polynomial fit to $K^{0}_{s}$ $v_{2}$
   and yellow band represent uncertainity due to fit.}
\end{figure}  
The observed NCQ
scaling for 0-80$\%$ collision centrality at RHIC was consider as a good signature
of partonic collectivity~\cite{rhicflow,mass}. It will be  interesting to investigate NCQ
scaling for different centrality as it could help us to understand
partonic collectivity for different system size. Figure 2 shows
$v_{2}$ scaled by number of constituent-quark ($ n_{q} $) as
function of $(m_{ T} -m_{0})/n_{q}$ in Au + Au collision at
$\sqrt{s_{NN}}$= 200 GeV for different collision centrality.
The ratio to the fit
line of the other hadron $v_{2}$ is shown in the corresponding lower
panels. 
Because of large statistical error, ratio for $\Omega$ is
not shown.
From the results in the panel (b) of figure 2 we find  that scaling holds for all identified strange
hadrons for 0-30$\%$ centrality.
 This indicates that the major part of flow could be developed at the
partonic phase for 0-30$\%$ centrality. On the other hand, for
30-80$\%$ centrality  shown in panel (c) of figure 2 we observe that
$\phi$-meson shows
a deviation of the order of 10$\%$ from the fit line  for the range
  $(m_{T} -m_{0})/n_{q}$  $>$ 0.6 GeV/$c^{2}$. This could be
  interpreted as due to  lower
  contribution from the partonic phase to the collectivity.

\begin{figure}[!ht]
\centerline{\includegraphics[scale=0.32]{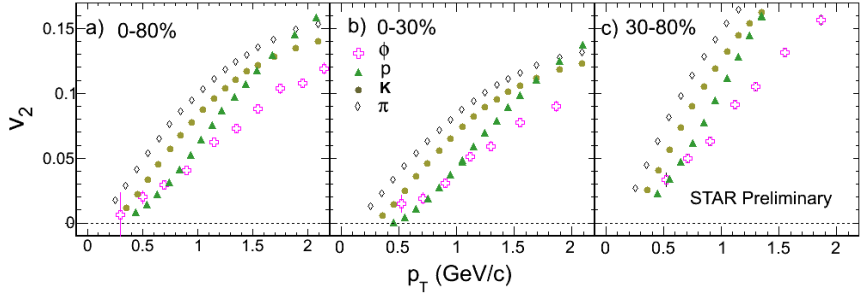}}
\caption{(Color online) $v_{2}$ vs. $p_{ T}$ of $\pi$, $\it{K}$, $\it{p}$ and $\phi$-meson for Au +
Au collision at $\sqrt{s_{NN}}$= 200 GeV with $|\it{y}| < $ 1.0 for centrality a)
   0-80 $\%$ b) 0-30 $\%$ and c) 30-80 $\%$. }
\end{figure} 
 
According to ideal hydrodynamics $v_{2} (p_{\mathrm T})$ follows a
mass ordering for $p_{T}$ $<$ 2 GeV/$c$,
For heavier  mass   $v_{2}$ is lower and vice-versa. In data, mass ordering was
observed in the low $p_{T}$ region~\cite{mass} . Recent phenomenological
calculation based on ideal hydrodynamical model together with the
hadron cascade shows that mass ordering of $v_{2}$ could be broken between that
$\phi$-meson  and proton at low $p_{ T}$ ($p_{T}$ $<$ 1.5 GeV/$c$
)~\cite{hyrdo_cascade}. This is because of late stage hadronic
re-scattering effects. High statistics data, collected in the year of
2010, allows for such an investigation. 
Figure 3 shows the comparison between $\pi$, $\it{K}$, $\it{p}$ and $\phi$-meson
$v_{2}$ for centrality 0-80$\%$, 0-30$\%$ and 30-80$\%$. $v_{2}$ of $K^{0}_{S}$,  $\Lambda$, $\Xi$ and $\Omega$
are not shown in figure 3 to make clear plot.
We observed mass ordering in $v_{2}$ for all identified particles like   $\pi$, $\it{K}$,
$K^{0}_{S}$, $\it{p}$, $\Lambda$, $\Xi$ and $\Omega$ except
$\phi$-meson. One can see from figure 3 for 0-30$\%$ centrality that at low $p_{ T} $ ($p_{ T}
$ $<$ 1 GeV/$c$) $\phi$-meson $v_{2}$ is either higher or similar to
that of proton $v_{2}$
although mass of $\phi$-meson (1.019 GeV/$c^{2}$) is greater than
mass of proton (0.938 GeV/$c^{2}$). This observation is consistent
with the scenario of
hadronic re-scattering effect as predicted in the theoretical
model~\cite{hyrdo_cascade}.
\section{Summary}
 
In summary, we present a systematic measurement of centrality dependence of
multi-strange hadrons $v_{2}$ at mid-rapidity using a high statistics data in Au + Au collisions at
$\sqrt{s_{NN}}$= 200 GeV collected in year 2010. We have observed a clear baryon-meson
splitting at intermediate $m_{ T} - m_{0}$ for centrality 0-30$\%$
and NCQ scaling hold for this centrality. This is consistent with the
idea of partonic collectivity.
The grouping among baryons and among mesons has been broken for 30-80$\%$ centrality.  Multi-Strange baryon (meson)  shows smaller $v_{2}$ than
that for other identified baryons (mesons) for 30-80$\%$
centrality. We observe $\phi$-meson $v_{2}$ shows a larger deviation in
quark-number scaling in 30-80$\%$ centrality than that for 0-30$\%$. It may indicate smaller contribution from the partonic phase to
the collectivity. In order to investigate effect of hadronic
re-scattering effect on $v_{2}$ , we have studied a comparison between
$\phi$-meson and proton $v_{2}$ in the low $p_{ T}$ region. We observe
$v_{2} ( \phi )$ $\geqslant$ $v_{2} (\it{p})$ for $p_{ T}$ $<$ 1.0 GeV/$c$ for  0-30$\%$ centrality. This observation is consistent with later stage
hadronic re-scattering effect as predicted in the theoretical model~\cite{hyrdo_cascade}.

\end{document}